\documentclass[reprint,aip]{revtex4-1} 

\usepackage{amsfonts,amssymb,amsmath,geometry, url,siunitx}
\usepackage{gensymb}
\usepackage{graphicx}
\usepackage{hyperref}
\usepackage{multirow}
\usepackage{setspace,upgreek}
\usepackage{stackengine}
\usepackage{subfig}
\usepackage{float}
\usepackage{collcell}
\usepackage{xstring}
\usepackage{datatool}
\usepackage{booktabs}
\usepackage{environ}
\usepackage{chngcntr}

 \hypersetup{
    colorlinks=true,    
    linkcolor=blue,       
    citecolor=red,       
    filecolor=red,        
    urlcolor=blue,        
    linktoc=page          
 }

\begin{document}
\title {Error Minimization in Predicting Accurate Adsorption Energies Using Machine Learning}
\author{Sanjay  Nayak}
\email{sanjay.kumar@ikst.res.in}
\affiliation{Indo-Korea Science and Technology Center, Bangalore-560064, India}
\author{Satadeep Bhattacharjee}
\email{satadeep.bhattacharjee@ikst.res.in}
\affiliation{Indo-Korea Science and Technology Center, Bangalore-560064, India}
\author{Jung-Hae Choi}
\affiliation{Center for Electronic Materials, Post-Silicon Semiconductor Institute, Korean Institute of Science and Technology, Seoul 02792, Republic of Korea}
\author{Seung Cheol Lee}
\email{seungcheol.lee@ikst.res.in}
\affiliation{Indo-Korea Science and Technology Center, Bangalore-560064, India}
 
\begin{abstract}

Finding the ``ideal" catalyst is a matter of great interest in the communities of chemists and material scientists, partly because of its wide spectrum of industrial applications. Information regarding a physical parameter termed ``adsorption energy", which dictates the degrees of adhesion of an adsorbate on a substrate is a primary requirement in selecting the catalyst for catalytic reactions.  Both experiments and \textit{in-silico} modelling are extensively being used in estimating the adsorption energies, both of which are \textit{Edisonian} approach and demands plenty of  resources and are time consuming. In this report, by employing a data centric approach almost instantly we predict the adsorption energies of atomic and molecular gases on the surfaces of many transition metals (TMs). With less than 10 sets of simple atomic features, our predictions of the adsorption energies are within a root-mean-squared-error (RMSE) of less than 0.4 eV with the quantum many-body perturbation theory estimates, a computationally expensive  with good experimental agreement. Further, we minimized the RMSE up to 0.11 eV by using the precomputed adsorption energies obtained with conventional exchange and correlation (XC) functional as one component of the feature vector. Based on our results, we developed a set of scaling laws between the adsorption energies computed with many-body perturbation theory and conventional DFT XC-functionals.

\end{abstract}

\maketitle


By reason of the ``Foundation pillar of green chemistry'', research in the field of catalytic and biocatalytic reactions has been going on from the last two century\cite{anastas2001catalysis}. The ultimate goal of these research is to find an ``ideal catalyst" which is thermally, mechanically and chemically stable, environmental friendly, posses good selectivity, and efficient in reducing the reaction barrier between reactants and product\cite{ananikov2012toward}.  The mechanism of catalytic reaction is a multi-step process. Especially, in the heterogeneous catalytic process, where catalysts are often surfaces of solids, majorly governed by three steps (i) adsorption of the reactants (i.e. molecules), (ii) hold the reactant in close proximity for the chemical reaction to take place and (iii) lets the product desorb back to the sorrounding. Being the first-stage of the reaction process, adsorption of molecules on catalyst surface have a large influences on the surface reaction\cite{norskov2008nature, hammer1995gold} which is visible in famous Br$\o$nsted-Evans-Polanyi (BEP) relationships\cite{bligaard2004bronsted}. Classic works of Christensen \textit{et al}\cite{norskov2008nature} and Sabatier \textit{et al}\cite{sabatier1911hydrogenations} further shows that optimal adsorption energies of reactants will maximize the catalytic activity. Thus, accurate determination of the adsorption energy is important and necessary in selecting suitable catalyst for the reactions process. 
\par
Both experiments and theoretical calculations were extensively used in past 
to determine the adsorption energies and reaction barriers of various adsorbate on  many solid sufaces\cite{janssens2007insights,katsanos1999measurement,rudzinski2012adsorption,liu2015quantitative,kibsgaard2015designing}. Ab-initio quantum mechanical simulation has emerged as a powerful tools not only in determing the adsorption energy but also in understanding the microscopic mechanism of catalytic reactions. However, estimates from these quantum mechanical simulations are functional dependent and the obtained results are varies widely with differnt functional. There are numbers of various methods are being developed and are discussed in the literature. Many of these methods are very selective in appropriate estimation to some specific properties only \cite{schmidt2018benchmark} out of which Random Phase Approximation (RPA)\cite{furche2001molecular,fuchs2002accurate,ren2012random,marini2006first}, a method beyond standard DFT is now considered to be a gold standard for solid state systems. Recently, Schmidt \textit{et al.} \cite{schmidt2018benchmark} benchmarked the RPA estimated adsorption energies to experimentally obtainbed values with an mean signed error (MSE) of $\approx$ -0.10 eV. Authors further shows that using standard DFT functionals the mean absolute error (MAE) in estimation of the adsorption energies are in the ranges of $\approx$ 0.20-0.44 eV \cite{schmidt2018benchmark}. However, such remarkable reduction in the error using RPA comes with an overload of computational cost. It is noteworthy to mention here that, calculations using standard DFT XC-functionals itself a time consuming process especially for a large number of systems.

\par

The emergence of contemporary Machine Learning (ML) technique which is a data centric approach to solve problems has shown a great promise in the field of material science too. Recently, many research groups employed this data centric approach to the field of catalysis\cite{ras2013predicting, takigawa2016machine, chowdhury2018prediction, li2017high, jager2018machine}. Rothenberg \textit{et al.}\cite{ras2013predicting} shows that with a simple set of feature vector the adsorption energies can be predicted by an RMSE  of 0.94-1.16 eV and  R$^2$ of 0.95; in the case of gas adsorbate on a range of metal surfaces. The RMSEs in this case are a bit higher. Similarly, Takigawa \textit{et al.} predicted the energies of d-band centers in  metals and bimetals, which is a crucial factor in determination of adsoption energy of gases on metal surfaces, with an RMSE of less than 0.5 eV. Recent work of Chowdhury \textit{et al.} shows that predicted adsorption energy using ML can be as close to the quantum mechanically calculated values with an MAE of 0.12 eV\cite{chowdhury2018prediction}, however the descriptors in this work are not straight forward.
\par
To address the above mentioned issues collectively in this letter, we show that with a suitable model and tuned hyperparameters the adsorption energies can be predictable quite accurately and instantly. More specifically, we predict the accurate adsorption energies of atomic (H, N, and O) and O-X (X=H,C, and N) molecular species on fcc (111) surfaces of 25 different TMs (Sc, Ti, V, Cr, Mn, Fe, Co, Ni, Cu, Y, Zr, Nb, Mo, Ru, Rh, Pd, Ag, Hf, Ta, W, Re, Os, Ir, Pt and Au) within the RMSE of ..computational error by employing an advanced supervised ML technique. We also provides a set of scaling laws to deduce accurate adsorption energies from estimates with relatively low accuracy.

In this work, we used the data, from an openly accesible database Computational Materials Repository (CMR),  generated by Schmidt \textit{et al.} \cite{schmidt2018benchmark}. Here, author used \textit{first-principles} DFT based calculations for the estimation of the adsorption energies and surface free energies with various XC-functional (\textit{e.g.} PBE, RPBE, BEEF-vdW, etc) along with RPA computed ones. For the estimation of adsorption energies, the adsorbate site was chosen to be at top site of fcc (111) surfaces. To reduce the computational cost, authors calculated the adsorption energies with relatively high surface coverages.

\par
We divided the dataset into two part; (i) 75\% for training and (ii) 25\% for testing cases. As feature vector we used majorly elemental properties of individual elements which constitute both surfaces and adsorbates. These elemental properties are groups (G), Covalent Radii (R), atomic number (AN), atomic mass (AM), period (P), Electronegativity (EN),  first ionization energy (IE), and Enthalpy of fusion ($\Delta_{fus}H$) (see Tab S-1 of suppplementary information (SI) for details). Structural properties such as density ($\rho$) of each species is also used. The molecules are further characterized by their HOMO and LUMO levels which are calculated using an STO-3G basis set and are obtained from NIST database and Ref.\onlinecite{zhang2007comparison}. The surfaces are further characterized by by their surface free energies ($\sigma$), Work function ($\Phi$)\cite{skriver1992surface} and Weigner seitz radius ($r_s$) of the respective elements  \cite{deb1992universal, takigawa2018machine}. Pearson's correlation between properties of elements are presented in the form of heatmap in Fig.S1 of SI.

To identify the best model for the prediction of adsorption energy of above mentioned gases on TMs surfaces, we employed various linear (\textit{e.g.} OLS, PLS, Ridge, Lasso and Kernel Ridge) and non-linear regression (\textit{e.g.} Gaussian Process (GPR), Gradient Boosting (GBR) and Random Forest Regression (RFR)) methods as implemented in \texttt{Scikit-Learn} package\cite{pedregosa2011scikit}. The predictibility of the models are assesed by Monte Carlo cross validation method. For comparative analysis between the RMSE in predictions by different regression model and brief discussions on methods of KRR, GBR and RFR see section II of SI. It is to mention that, prior to all ML regression we standardize the data by transforming them to center it by removing the mean value of each feature, then scale it by dividing non-constant features by their standard deviation, for which we used the \texttt{preprocessing} tools of \texttt{Scikit-Learn} package. 
\par
To this end, we arrange rest of the texts in following order;

\begin{enumerate}
	\item First we discuss the ML prediction of adsorption energies using a simple set of feature vector
	\item Next we emphasizes on the substantial reduction in errors in prediction of accurate adsorption energies by using less accurately (with low computational cost) precomputed adsorption energies as an addition to the set of previous feature vector
\end{enumerate}

\begin{figure*}[!ht]
	\centering
	\includegraphics[scale=1.3]{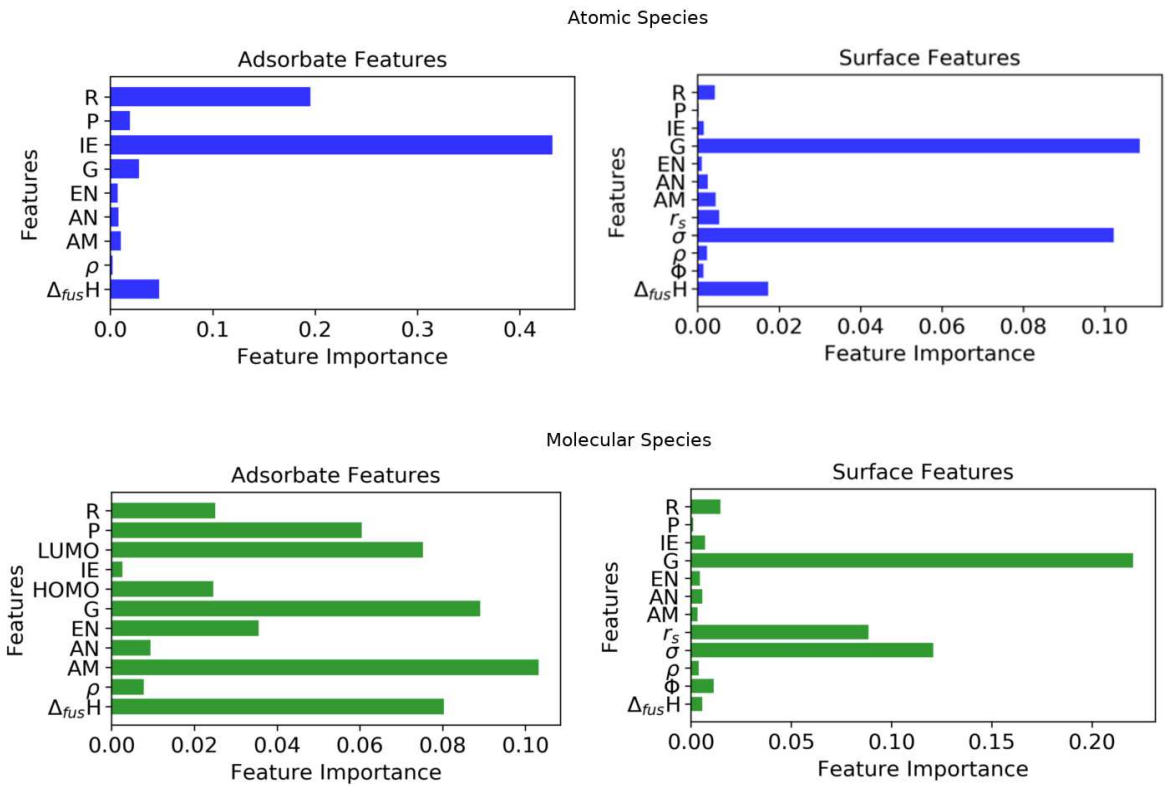}
	\caption{Feature importance obtained with Gradient Boosting Regression (GBR). }
	\label{FI}
\end{figure*}



In this work, we treated both atomic and molecular adsorbate separately.
We started our analysis in prediction of adsorption energies of atomic gases (H,C and N), on the surfaces of 25 TMs with a total 21-dimensional feature vector. We employed various regression methods, and compare their RMSEs of predictions (see Fig. S2 of SI). It is to note that the mentioned (here and afterwards) RMSEs and SDs are obtained with Monte-carlo cross-validation method. While most linear regressions method can predict the adsorption energies within an RMSE of 0.86 eV and a standard deviations (SD) of $\approx$ 0.11 eV, KRR predicts an RMSE of 0.35 eV with SD of 0.06 eV. Surprisingly linear regression model like KRR predicts with better accuaracies than non-linear regression techniques GPR, GBR and RFR where RMSE (SD) of 0.50(0.27), 0.43(0.13) and 0.50(0.13) eV are noted respectively. The R$^2$ score of the training set in KRR, GBR and RFR is 0.990, 0.999, and 0.987  while for testing set it is 0.95, 0.93 and 0.91, respectively. Such good fitting of data in both training and testing cases indicate that these models neither overfitting or underfitting (see Fig. S2 and Fig. S3(a-c)).
\par

\begin{figure*} [!ht]
	\centering
	\subfloat[ ]{{\includegraphics[scale=0.65]{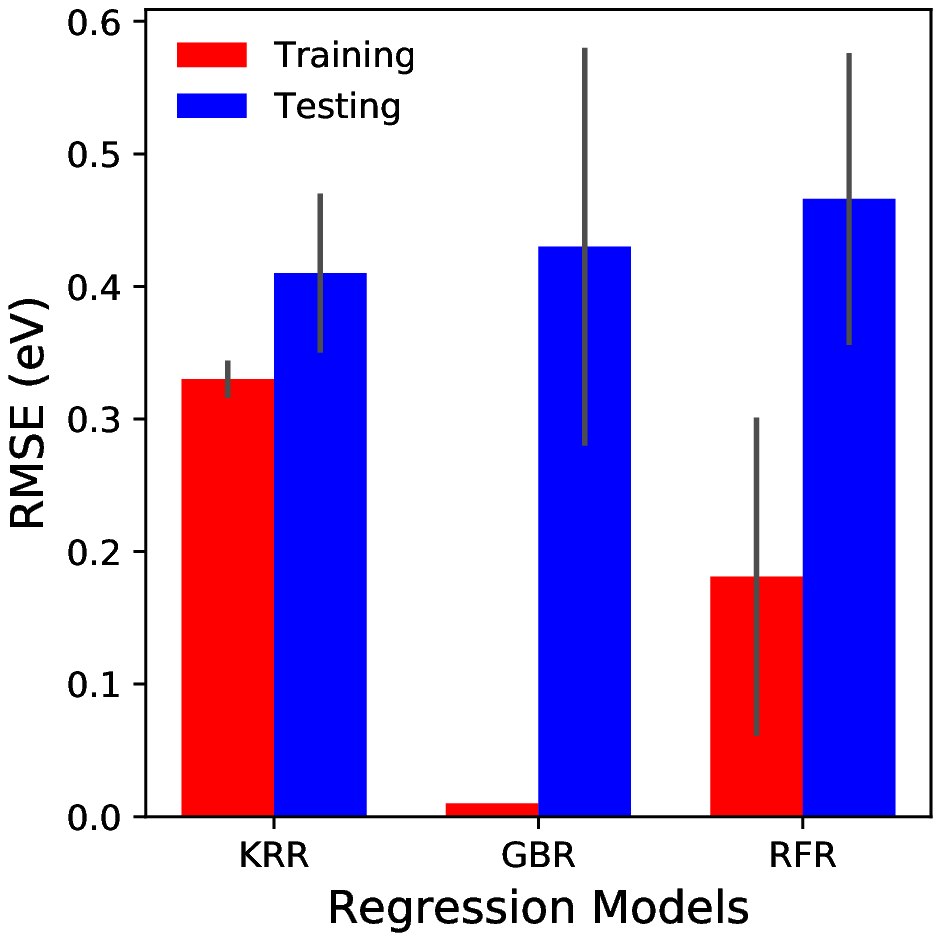} }}
	\subfloat[ ]{{\includegraphics[scale=0.65]{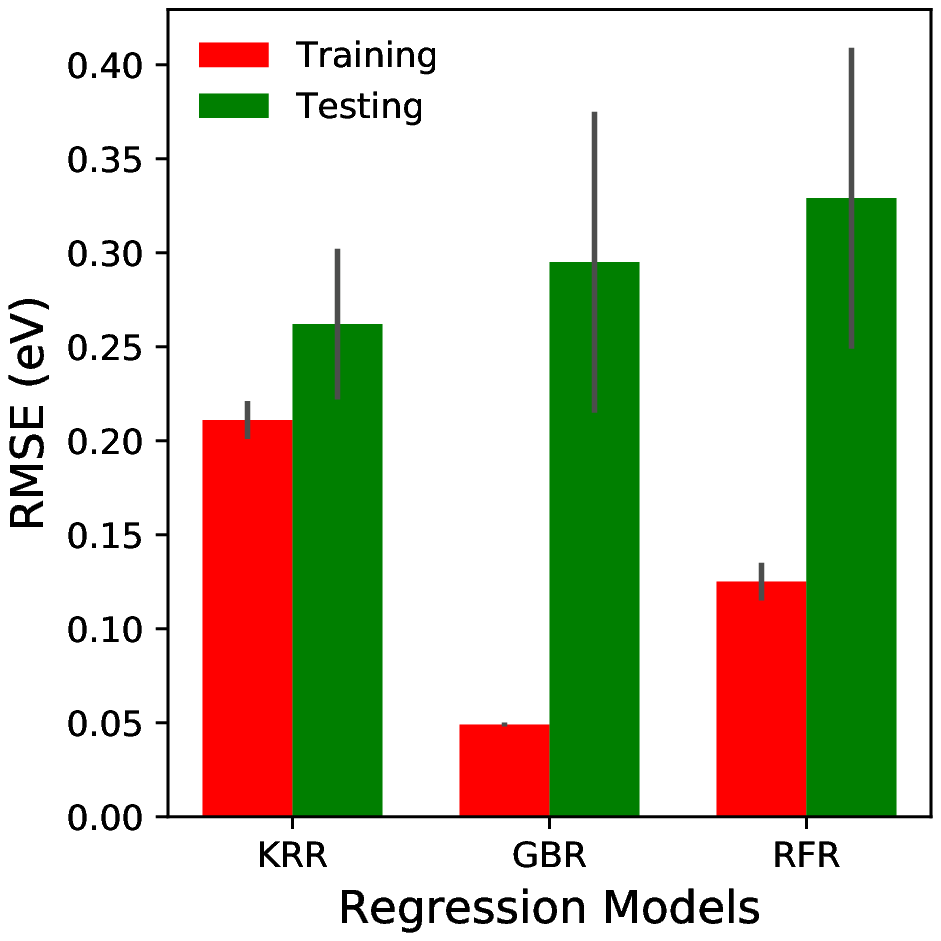} }}%
	\caption{RMSE in training and testing cases for atomic (a) and molecular adsorbates(b) where reduced features are used.}%
	\label{red_error}%
\end{figure*}

To minimize the number of features, we obtained their importance in prediction of adsorption energy from GBR method (see Fig. \ref{FI}). Features like covalent radius (R), first ionization energy (IE) of adsorbate and enthalpy of fusion ($\Delta_{fus}H$) at 300 K of adsorbates are important for adsorbate, while for metal surfaces features like Group (G), surface free energies ($\sigma$), Wigner-Seitz radius ($r_s$) and enthalpy of fusion ($\Delta_{fus}H$) at 300 K are important. Thus we construct a reduced 7-dimensional feature vector with the above mentioned ones. We find that statistically obtained results with these reduced features are as good as obtained with the earlier 21-dimensional features. The training RMSE (SD) obtained with KRR, GBR and RFR are 0.33 (0.01), 0.01 (0.00), 0.18 (0.12) eV while for testing these are 0.41(0.06),0.43(0.15) and 0.46(0.11) eV respectively (see Fig.\ref{red_error} (a) and Fig. S4 (a-c) for fitting). 

 Similar to atomic species,  we started our prediction of adsorption energies of O-X molecules with a total of 23-dimensional feature vector. Again in this case KRR out-perform other methods considered here (see Fig.S2 of SI). The predicted RMSE (SD) of test set with KRR, GBR and RFR are 0.27(0.05), 0.32(0.09), 0.38(0.09) eV respectively (see Fig.S2 and S3(d-f) of SI). We note that besides these three methods, GPR works relatively well in molecular systems with a prediction RMSE and SD of 0.28 and 0.13 eV respectively. For O-X molecules obtained feature importance from GBR revealed that LUMO levels of O-X molecules, Period (P),  Group (G), atomic mass (AM) and enthalpy of fusion ($\Delta_{fus}H$) of X (=H,C, and N) while Group (G), Wigner Seitz radius ($r_s$) and surface free energies ($\sigma$) of metal surface is primarily determining the adsorption energies. Based on these observations, we constructed a 8-dimensional reduced descriptors and predict the adsorption energy again using ML. The RMSEs (SD) in predicting the adsorption energies with reduced set of descriptors for training data set with KRR, GBR, and RFR are 0.21(0.01), 0.05(0.00) and 0.12(0.01) eV respectively while for testing cases the obtained RMSEs (SD) are 0.26(0.04), 0.29 (0.08), 0.33 (0.08) eV respectively (see Fig.\ref{red_error}(b)). 
 
 \par
 Thus, our work suggest that using ML technique with a simple set of feature vector, adsorption energies of atomic (O, H, and N) and molecular species (O-X) on TM surfaces can be predictable upto an RMSE of 0.41 eV and 0.26 eV respectively. To further minimize this error, we used  a set of precomputed DFT adsorption energies as a component of feature vector. Next part of this paper is dedicated to the same.

\begin{figure}[!h]
	\centering
	\includegraphics[scale=0.45]{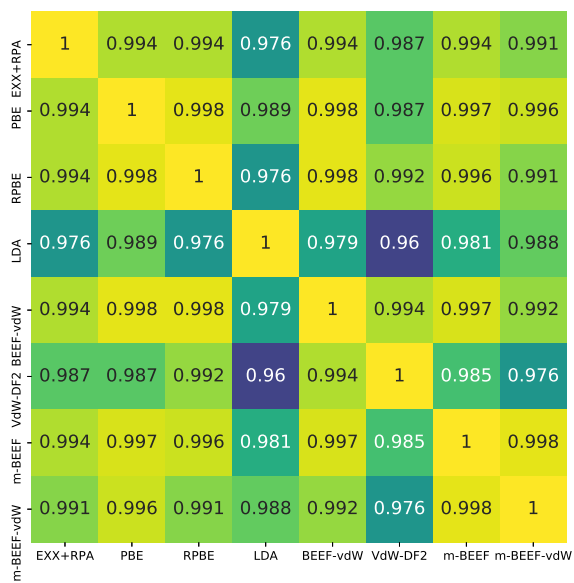}
	\caption{Correlation map of Pearson's correlation coefficient ($p$) of adsorption energies obtained with various DFT XC-functionals and RPA.}
	\label{func-correlation}
\end{figure}

\begin{figure}[!hb]
	\centering
	\subfloat[ ]{{\includegraphics[width=6cm]{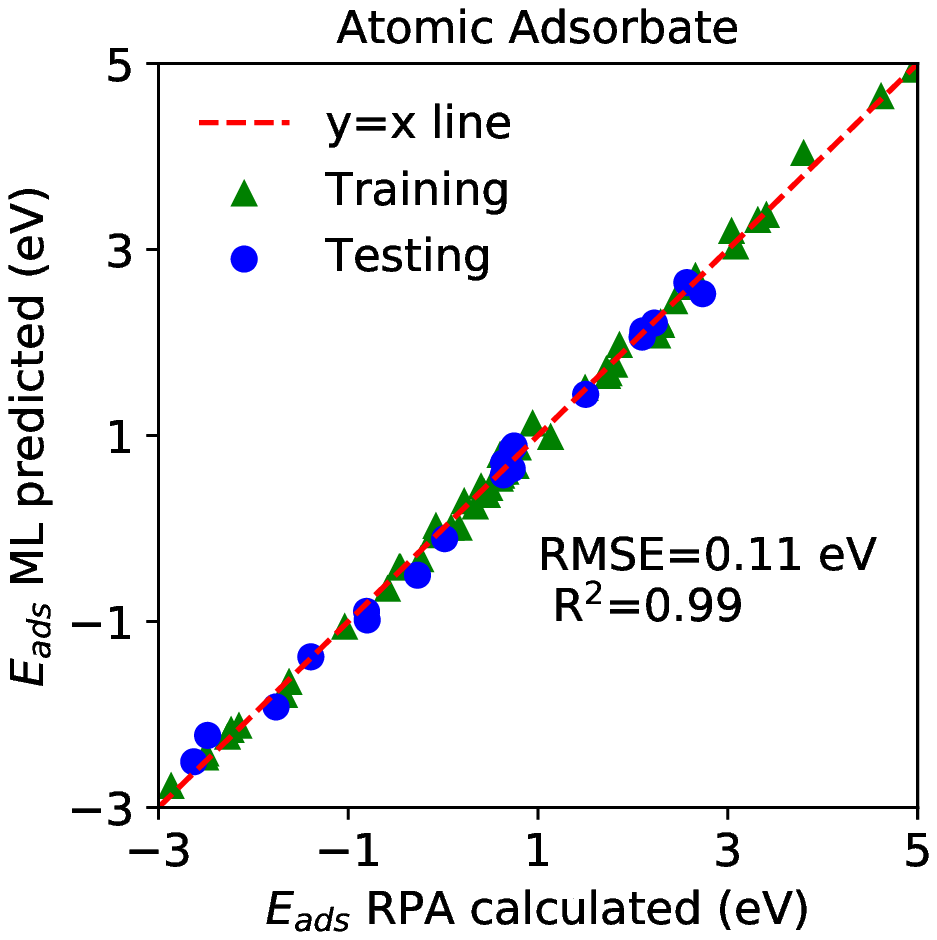} }}%
	\
	\subfloat[ ]{{\includegraphics[width=6cm]{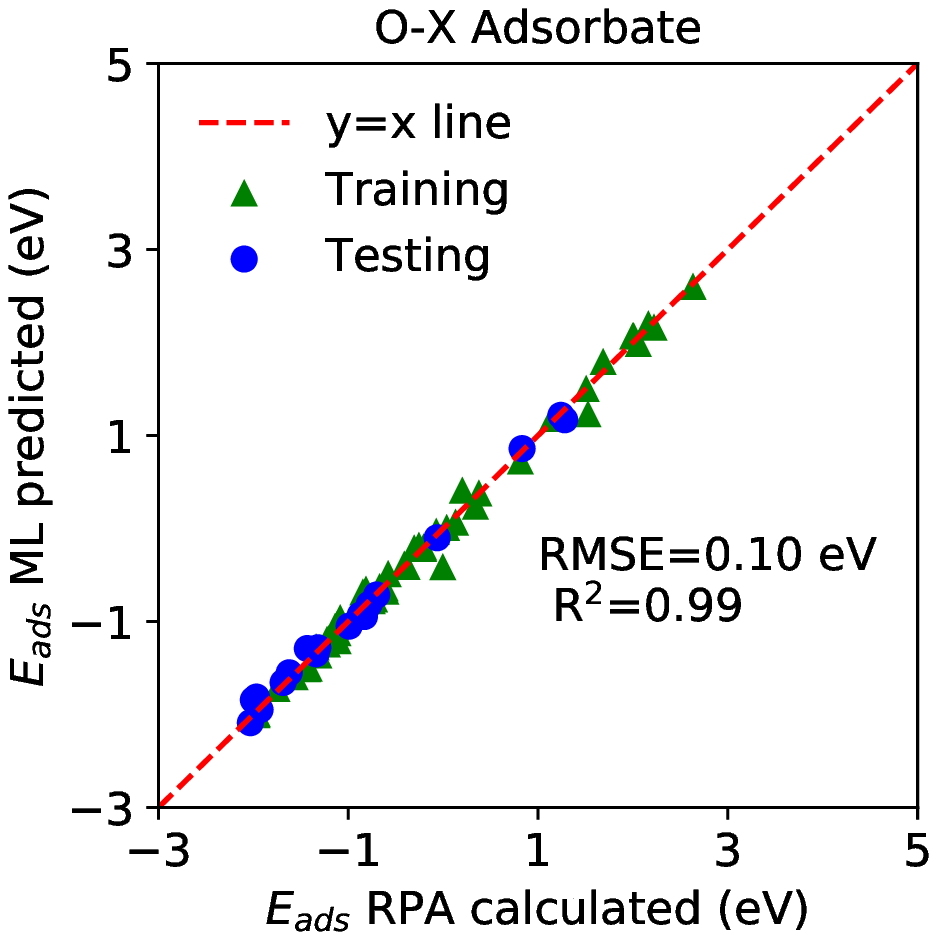} }}%
	\caption{(a)  and (b) show adsorption energies predicted using OLS regression methods for Atomic and molecular adsorbate, respectively. Note that in these predictions PBE computed adsorption energies are also used as one of the descriptors. For details see main text.}%
	\label{final}%
\end{figure}

We obtained  Pearson's correlation between adsorption energies estimated with various XC-functional which shows that they are highly correlated ($p>0.96$) to each other as well as to RPA estimated ones (see Fig.\ref{func-correlation}). This observation further motivated us to find a set of scaling laws between RPA estimated adsorption energies and less accurately computed with standard DFT XC-functionals, as RPA method required a huge computational resources. Now alongwith the precomputed adsorption energies using standard DFT XC-functionals
we also used the important features that we obtained from previous observations of this work.

\begin{table*}
	\caption{Weight coefficient obtained with ordinary least square (OLS) regression for atomic adsorbate.} \label{tab2}
	\begin{tabular}{|c|c|c|c|c|c|c|c|c| lllllllll@{}}
		\hline
		& RPA & PBE    & RPBE   & LDA    & BEEF-vdW & vdW-DF2 & m-BEEF & m-BEEF-vdW \\
		\hline
		E$_{ads.}$             & 1   & 1.635  & 1.672  & 1.540  & 1.710    & 1.826   & 1.598  & 1.582      \\ \hline
		R(ad)                     & 0   & -0.113 & -0.118 & -0.08  & -0.070   & 0.063   & -0.151 & -0.150     \\ \hline
		IE(ad)                   & 0   & 0.180  & 0.133  & 0.259  & 0.102    & -0.060  & 0.221  & 0.232      \\ \hline
		$\Delta_{fus}H$(ad)  & 0   & 0.019  & -0.008 & 0.076  & 0.007    & 0.101   & 0.106  & 0.022      \\ \hline
		G (surf.)                    & 0   & -0.117 & -0.118 & -0.053 & -0.175   & -0.246  & -0.072 & -0.051     \\ \hline
		$\Delta_{fus}H$(surf.) & 0   & -0.087 & -0.081 & -0.137 & -0.097   & -0.133  & -0.074 & -0.087     \\ \hline
		$r_s$ (surf.)                 & 0   & -0.104 & -0.071 & -0.129 & -0.101   & -0.087  & -0.024 & -0.053     \\ \hline
		$\sigma$ (surf.)            & 0   & 0.028  & 0.023  & 0.091  & 0.044    & 0.016   & 0.063  & 0.089     \\
		\hline
	\end{tabular}
\end{table*}

\begin{table*}
	\centering
	\caption{Weight coefficient obtained with ordinary least square (OLS) regression for O-X adsorbate.} \label{tab3}
	\begin{tabular}{|c|c|c|c|c|c|c|c|c|c| lllllllll@{}} 
		\hline
		& RPA & PBE    & RPBE   & LDA    & BEEF-vdW & vdW-DF2 & m-BEEF & m-BEEF-vdW \\ \hline
		E$_{ads.}$          & 1   & 1.369  & 1.395  & 1.297  & 1.377    & 1.336   & 1.243  & 1.251      \\ \hline
		G(X)               & 0   & 0.102  & 0.105  & 0.118  & 0.086    & 0.097   & 0.019  & 0.012      \\ \hline
		AM(X)              & 0   & 0.077  & 0.070  & 0.930  & 0.055    & 0.039   & 0.020  & 0.027      \\ \hline
		$\Delta _{fus} H$(X) & 0   & -0.047 & -0.073 & -0.044 & -0.065   & 0.136   & 0.005  & 0.041      \\ \hline
		P(X)               & 0   & 0.064  & 0.054  & 0.079  & 0.041    & 0.015   & 0.019  & 0.032      \\ \hline
		G (surf.)                  & 0   & -0.145 & -0.155 & -0.056 & -0.209   & -0.375  & -0.046 & -0.059     \\ \hline
		$r_s$ (surf.)             & 0   & -0.076 & 0.027  & -0.264 & -0.013   & -0.008  & -0.023 & -0.108     \\ \hline
		$\sigma$ (surf.)          & 0   & -0.030 & -0.028 & -0.039 & -0.043   & -0.184  & 0.024  & 0.031      \\ \hline
		$E_{LUMO}$ (O-X)               & 0   & -0.015 & 0.004  & -0.027 & -0.007   & 0.062   & -0.015 & -0.041     \\ \hline 
	\end{tabular}
\end{table*}

For the prediction of accurate adsorption energies of atomic adsorbate on various TMs surface an 8-dimensional feature vector is used. Surprisingly, a simple ordinary least squares (OLS) regression method predicts the adsorption energies within an RMSE $<$ 0.15 eV which is more than 130\% reduction in the error compared to the previous case where conventional DFT XC-functional's estimates is not used as a feature. For details of the error in estimating accurate adsorption energies with various DFT XC-functional estimated adsorption energies see Fig.S5 of SI. In Fig.\ref{final}(a) we presented adsorption energies with RPA and our ML predicted where PBE estimated adsorption energies are used as one of the feature vector. An RMSE in the prediction is noted to 0.11 eV with a SD of 0.017 eV.  It is important to note here that on average the adsorption energies  computed with RPA deviates by $\approx$ 0.2 eV from experimentally determined values. Thus, our ML predicted adsorption energies are within the small differences of RPA computed and experimentally obtained values. Since an OLS regression predicts  the adsorption energies quite accurately, we deduce a set of scaling laws based on the obtained weight coefficient of features which takes a general form of:
 
 \begin{equation}
 E_{ads}^{RPA} = \sum_{i} \alpha_i D_i
 \end{equation}
 where $\alpha_i$s are the weight co-efficients of the individual features $D_i$ for a particular XC-functional and are tabulated in Table \ref{tab2}. For example, RPA estimated adsorption energy can be obtained from PBE estimated by using the following relations;
 
 \begin{equation*}
  \begin{aligned}
 {} & E_{ads}^{RPA} (atm.\ sp.) = 1.635 \times E_{ads}^{PBE}-0.113\times R (ad) \\
 & \quad \quad + 0.180 \times IE (ad)  + 0.019 \times \Delta_{fus.}H(ad) \\
 & \quad \quad - 0.117 \times G (surf.)   - 0.087 \times \Delta_{fus.} H(surf.)\\  & \quad \quad -0.104 \times r_s (surf.)  + 0.028\times \sigma^{PBE}
  \end{aligned}
 \end{equation*}
  Similar to PBE, scaling between RPA and other XC-functionals can be written using information given in Table \ref{tab2}.

Further, we scaled down the adsorption energies obtained with different XC-functional to RPA estimates for O-X molecules too. Here we used a 9-dimensional feature vector with which OLS regression predicts adsorption energies within an RMSE of less than 0.11 eV except for LDA and vdW-DF2 functional. Obtained RMSEs of LDA and vdW-DF2 functionals are 0.19 and 0.16 eV respectively. Our prediction of RPA adsorption energies using PBE estimates as a feature has an RMSE of 0.10 eV with a SD of 0.02 eV (see Fig.\ref{final}(b) and  Fig.S5(b) of SI). Here we again find that the reduction in the RMSE value is  $\approx$ 130\% compare to the previous estimate discussed earlier. Further, relation between RPA and DFT XC-functional based estimate takes a general form of;

\begin{equation}
E_{ads}^{RPA}=\sum_{i} \beta_i D_i
\end{equation} 
where $\beta_i$ has the similar meaning of $\alpha_i$ and values are given in Table \ref{tab3}. Scaling relation between RPA and DFT-PBE  estimate is given by the following equation;
\begin{equation*}
\begin{aligned}
{} & E_{ads}^{RPA} (O-X) = 1.369 \times E_{ads}^{PBE}+0.102\times G(X) \\ &\quad \quad + 0.077 \times AM(X)  -0.047\times \Delta_{fus.}H(ad)  \\ & \quad \quad + 0.064 \times P(X) -0.145 \times G
  - 0.076 \times r_s (surf.) \\ & \quad \quad -0.030 \times \sigma^{PBE} (surf.)-0.015 \times E_{LUMO} (O-X)
\end{aligned}
\end{equation*}

\par
Thus, here we show that error (RMSE) in predictions of adsorption energies of atomic and O-X gas adsorbates on many transition metal surfaces can be reduced to the level of the  difference between the estimates of computationally powerful many body perturbation theory and the experimentally obtained ones, using a state-of-the-art supervised machine learning tools. We presented a set of scaling laws which can be used to correct the adsorption energies obtained with low accuracy and relatively inexpensive computation (such as DFT-PBE) to the computationally expensive and high accurately computed (RPA) ones. We also suggest here that  proposed method to reduce the error in predictions of accurate physical properties need not to be necessarily valid only for adsorption energy but should also be tested for  other material properties.
 
In summary, we predict adsorption energies of H, O, N, O-H, N-O and C-O on fcc (111) surfaces of 25 different elemental transition metals using state-of-the-art supervised machine learning technique. With only atomic informations as features, we predict the accurate adsorption energies within an RMSE of $<$ 0.4 eV. By using a set of pre-computed adsorption energies  by conventional exchange and correlation functional, the RMSE in predictions reduced to the level of the error in high level quantum many body perturbation theory estimates as well as experimental methods. In general, we proposed a new method to obtain material properties with high accuracy, obtained with expensive computational cost, from a less accurately precomputed data obtained with relatively inexpensive computational methods.

\bibliographystyle{apsrev4-1}
\bibliography{bib.bib}

\begin{thebibliography}{29}%
\makeatletter
\providecommand \@ifxundefined [1]{%
 \@ifx{#1\undefined}
}%
\providecommand \@ifnum [1]{%
 \ifnum #1\expandafter \@firstoftwo
 \else \expandafter \@secondoftwo
 \fi
}%
\providecommand \@ifx [1]{%
 \ifx #1\expandafter \@firstoftwo
 \else \expandafter \@secondoftwo
 \fi
}%
\providecommand \natexlab [1]{#1}%
\providecommand \enquote  [1]{``#1''}%
\providecommand \bibnamefont  [1]{#1}%
\providecommand \bibfnamefont [1]{#1}%
\providecommand \citenamefont [1]{#1}%
\providecommand \href@noop [0]{\@secondoftwo}%
\providecommand \href [0]{\begingroup \@sanitize@url \@href}%
\providecommand \@href[1]{\@@startlink{#1}\@@href}%
\providecommand \@@href[1]{\endgroup#1\@@endlink}%
\providecommand \@sanitize@url [0]{\catcode `\\12\catcode `\$12\catcode
  `\&12\catcode `\#12\catcode `\^12\catcode `\_12\catcode `\%12\relax}%
\providecommand \@@startlink[1]{}%
\providecommand \@@endlink[0]{}%
\providecommand \url  [0]{\begingroup\@sanitize@url \@url }%
\providecommand \@url [1]{\endgroup\@href {#1}{\urlprefix }}%
\providecommand \urlprefix  [0]{URL }%
\providecommand \Eprint [0]{\href }%
\providecommand \doibase [0]{http://dx.doi.org/}%
\providecommand \selectlanguage [0]{\@gobble}%
\providecommand \bibinfo  [0]{\@secondoftwo}%
\providecommand \bibfield  [0]{\@secondoftwo}%
\providecommand \translation [1]{[#1]}%
\providecommand \BibitemOpen [0]{}%
\providecommand \bibitemStop [0]{}%
\providecommand \bibitemNoStop [0]{.\EOS\space}%
\providecommand \EOS [0]{\spacefactor3000\relax}%
\providecommand \BibitemShut  [1]{\csname bibitem#1\endcsname}%
\let\auto@bib@innerbib\@empty
\bibitem [{\citenamefont {Anastas}\ \emph {et~al.}(2001)\citenamefont
  {Anastas}, \citenamefont {Kirchhoff},\ and\ \citenamefont
  {Williamson}}]{anastas2001catalysis}%
  \BibitemOpen
  \bibfield  {author} {\bibinfo {author} {\bibfnamefont {P.~T.}\ \bibnamefont
  {Anastas}}, \bibinfo {author} {\bibfnamefont {M.~M.}\ \bibnamefont
  {Kirchhoff}}, \ and\ \bibinfo {author} {\bibfnamefont {T.~C.}\ \bibnamefont
  {Williamson}},\ }\href@noop {} {\bibfield  {journal} {\bibinfo  {journal}
  {Applied Catalysis A: General}\ }\textbf {\bibinfo {volume} {221}},\ \bibinfo
  {pages} {3} (\bibinfo {year} {2001})}\BibitemShut {NoStop}%
\bibitem [{\citenamefont {Ananikov}\ and\ \citenamefont
  {Beletskaya}(2012)}]{ananikov2012toward}%
  \BibitemOpen
  \bibfield  {author} {\bibinfo {author} {\bibfnamefont {V.~P.}\ \bibnamefont
  {Ananikov}}\ and\ \bibinfo {author} {\bibfnamefont {I.~P.}\ \bibnamefont
  {Beletskaya}},\ }\href@noop {} {\bibfield  {journal} {\bibinfo  {journal}
  {Organometallics}\ }\textbf {\bibinfo {volume} {31}},\ \bibinfo {pages}
  {1595} (\bibinfo {year} {2012})}\BibitemShut {NoStop}%
\bibitem [{\citenamefont {N{\o}rskov}\ \emph {et~al.}(2008)\citenamefont
  {N{\o}rskov}, \citenamefont {Bligaard}, \citenamefont {Hvolb{\ae}k},
  \citenamefont {Abild-Pedersen}, \citenamefont {Chorkendorff},\ and\
  \citenamefont {Christensen}}]{norskov2008nature}%
  \BibitemOpen
  \bibfield  {author} {\bibinfo {author} {\bibfnamefont {J.~K.}\ \bibnamefont
  {N{\o}rskov}}, \bibinfo {author} {\bibfnamefont {T.}~\bibnamefont
  {Bligaard}}, \bibinfo {author} {\bibfnamefont {B.}~\bibnamefont
  {Hvolb{\ae}k}}, \bibinfo {author} {\bibfnamefont {F.}~\bibnamefont
  {Abild-Pedersen}}, \bibinfo {author} {\bibfnamefont {I.}~\bibnamefont
  {Chorkendorff}}, \ and\ \bibinfo {author} {\bibfnamefont {C.~H.}\
  \bibnamefont {Christensen}},\ }\href@noop {} {\bibfield  {journal} {\bibinfo
  {journal} {Chemical Society Reviews}\ }\textbf {\bibinfo {volume} {37}},\
  \bibinfo {pages} {2163} (\bibinfo {year} {2008})}\BibitemShut {NoStop}%
\bibitem [{\citenamefont {Hammer}\ and\ \citenamefont
  {Norskov}(1995)}]{hammer1995gold}%
  \BibitemOpen
  \bibfield  {author} {\bibinfo {author} {\bibfnamefont {B.}~\bibnamefont
  {Hammer}}\ and\ \bibinfo {author} {\bibfnamefont {J.}~\bibnamefont
  {Norskov}},\ }\href@noop {} {\bibfield  {journal} {\bibinfo  {journal}
  {Nature}\ }\textbf {\bibinfo {volume} {376}},\ \bibinfo {pages} {238}
  (\bibinfo {year} {1995})}\BibitemShut {NoStop}%
\bibitem [{\citenamefont {Bligaard}\ \emph {et~al.}(2004)\citenamefont
  {Bligaard}, \citenamefont {N{\o}rskov}, \citenamefont {Dahl}, \citenamefont
  {Matthiesen}, \citenamefont {Christensen},\ and\ \citenamefont
  {Sehested}}]{bligaard2004bronsted}%
  \BibitemOpen
  \bibfield  {author} {\bibinfo {author} {\bibfnamefont {T.}~\bibnamefont
  {Bligaard}}, \bibinfo {author} {\bibfnamefont {J.~K.}\ \bibnamefont
  {N{\o}rskov}}, \bibinfo {author} {\bibfnamefont {S.}~\bibnamefont {Dahl}},
  \bibinfo {author} {\bibfnamefont {J.}~\bibnamefont {Matthiesen}}, \bibinfo
  {author} {\bibfnamefont {C.~H.}\ \bibnamefont {Christensen}}, \ and\ \bibinfo
  {author} {\bibfnamefont {J.}~\bibnamefont {Sehested}},\ }\href@noop {}
  {\bibfield  {journal} {\bibinfo  {journal} {Journal of Catalysis}\ }\textbf
  {\bibinfo {volume} {224}},\ \bibinfo {pages} {206} (\bibinfo {year}
  {2004})}\BibitemShut {NoStop}%
\bibitem [{\citenamefont {Sabatier}(1911)}]{sabatier1911hydrogenations}%
  \BibitemOpen
  \bibfield  {author} {\bibinfo {author} {\bibfnamefont {P.}~\bibnamefont
  {Sabatier}},\ }\href@noop {} {\bibfield  {journal} {\bibinfo  {journal}
  {Berichte der deutschen chemischen Gesellschaft}\ }\textbf {\bibinfo {volume}
  {44}},\ \bibinfo {pages} {1984} (\bibinfo {year} {1911})}\BibitemShut
  {NoStop}%
\bibitem [{\citenamefont {Janssens}\ \emph {et~al.}(2007)\citenamefont
  {Janssens}, \citenamefont {Clausen}, \citenamefont {Hvolb{\ae}k},
  \citenamefont {Falsig}, \citenamefont {Christensen}, \citenamefont
  {Bligaard},\ and\ \citenamefont {N{\o}rskov}}]{janssens2007insights}%
  \BibitemOpen
  \bibfield  {author} {\bibinfo {author} {\bibfnamefont {T.~V.}\ \bibnamefont
  {Janssens}}, \bibinfo {author} {\bibfnamefont {B.~S.}\ \bibnamefont
  {Clausen}}, \bibinfo {author} {\bibfnamefont {B.}~\bibnamefont
  {Hvolb{\ae}k}}, \bibinfo {author} {\bibfnamefont {H.}~\bibnamefont {Falsig}},
  \bibinfo {author} {\bibfnamefont {C.~H.}\ \bibnamefont {Christensen}},
  \bibinfo {author} {\bibfnamefont {T.}~\bibnamefont {Bligaard}}, \ and\
  \bibinfo {author} {\bibfnamefont {J.~K.}\ \bibnamefont {N{\o}rskov}},\
  }\href@noop {} {\bibfield  {journal} {\bibinfo  {journal} {Topics in
  Catalysis}\ }\textbf {\bibinfo {volume} {44}},\ \bibinfo {pages} {15}
  (\bibinfo {year} {2007})}\BibitemShut {NoStop}%
\bibitem [{\citenamefont {Katsanos}\ \emph {et~al.}(1999)\citenamefont
  {Katsanos}, \citenamefont {Rakintzis}, \citenamefont
  {Roubani-Kalantzopoulou}, \citenamefont {Arvanitopoulou},\ and\ \citenamefont
  {Kalantzopoulos}}]{katsanos1999measurement}%
  \BibitemOpen
  \bibfield  {author} {\bibinfo {author} {\bibfnamefont {N.~A.}\ \bibnamefont
  {Katsanos}}, \bibinfo {author} {\bibfnamefont {N.}~\bibnamefont {Rakintzis}},
  \bibinfo {author} {\bibfnamefont {F.}~\bibnamefont {Roubani-Kalantzopoulou}},
  \bibinfo {author} {\bibfnamefont {E.}~\bibnamefont {Arvanitopoulou}}, \ and\
  \bibinfo {author} {\bibfnamefont {A.}~\bibnamefont {Kalantzopoulos}},\
  }\href@noop {} {\bibfield  {journal} {\bibinfo  {journal} {Journal of
  Chromatography A}\ }\textbf {\bibinfo {volume} {845}},\ \bibinfo {pages}
  {103} (\bibinfo {year} {1999})}\BibitemShut {NoStop}%
\bibitem [{\citenamefont {Rudzinski}\ and\ \citenamefont
  {Everett}(2012)}]{rudzinski2012adsorption}%
  \BibitemOpen
  \bibfield  {author} {\bibinfo {author} {\bibfnamefont {W.}~\bibnamefont
  {Rudzinski}}\ and\ \bibinfo {author} {\bibfnamefont {D.~H.}\ \bibnamefont
  {Everett}},\ }\href@noop {} {\emph {\bibinfo {title} {Adsorption of gases on
  heterogeneous surfaces}}}\ (\bibinfo  {publisher} {Academic Press},\ \bibinfo
  {year} {2012})\BibitemShut {NoStop}%
\bibitem [{\citenamefont {Liu}\ \emph {et~al.}(2015)\citenamefont {Liu},
  \citenamefont {Maa{\ss}}, \citenamefont {Willenbockel}, \citenamefont
  {Bronner}, \citenamefont {Schulze}, \citenamefont {Soubatch}, \citenamefont
  {Tautz}, \citenamefont {Tegeder},\ and\ \citenamefont
  {Tkatchenko}}]{liu2015quantitative}%
  \BibitemOpen
  \bibfield  {author} {\bibinfo {author} {\bibfnamefont {W.}~\bibnamefont
  {Liu}}, \bibinfo {author} {\bibfnamefont {F.}~\bibnamefont {Maa{\ss}}},
  \bibinfo {author} {\bibfnamefont {M.}~\bibnamefont {Willenbockel}}, \bibinfo
  {author} {\bibfnamefont {C.}~\bibnamefont {Bronner}}, \bibinfo {author}
  {\bibfnamefont {M.}~\bibnamefont {Schulze}}, \bibinfo {author} {\bibfnamefont
  {S.}~\bibnamefont {Soubatch}}, \bibinfo {author} {\bibfnamefont {F.~S.}\
  \bibnamefont {Tautz}}, \bibinfo {author} {\bibfnamefont {P.}~\bibnamefont
  {Tegeder}}, \ and\ \bibinfo {author} {\bibfnamefont {A.}~\bibnamefont
  {Tkatchenko}},\ }\href@noop {} {\bibfield  {journal} {\bibinfo  {journal}
  {Physical review letters}\ }\textbf {\bibinfo {volume} {115}},\ \bibinfo
  {pages} {036104} (\bibinfo {year} {2015})}\BibitemShut {NoStop}%
\bibitem [{\citenamefont {Kibsgaard}\ \emph {et~al.}(2015)\citenamefont
  {Kibsgaard}, \citenamefont {Tsai}, \citenamefont {Chan}, \citenamefont
  {Benck}, \citenamefont {N{\o}rskov}, \citenamefont {Abild-Pedersen},\ and\
  \citenamefont {Jaramillo}}]{kibsgaard2015designing}%
  \BibitemOpen
  \bibfield  {author} {\bibinfo {author} {\bibfnamefont {J.}~\bibnamefont
  {Kibsgaard}}, \bibinfo {author} {\bibfnamefont {C.}~\bibnamefont {Tsai}},
  \bibinfo {author} {\bibfnamefont {K.}~\bibnamefont {Chan}}, \bibinfo {author}
  {\bibfnamefont {J.~D.}\ \bibnamefont {Benck}}, \bibinfo {author}
  {\bibfnamefont {J.~K.}\ \bibnamefont {N{\o}rskov}}, \bibinfo {author}
  {\bibfnamefont {F.}~\bibnamefont {Abild-Pedersen}}, \ and\ \bibinfo {author}
  {\bibfnamefont {T.~F.}\ \bibnamefont {Jaramillo}},\ }\href@noop {} {\bibfield
   {journal} {\bibinfo  {journal} {Energy \& Environmental Science}\ }\textbf
  {\bibinfo {volume} {8}},\ \bibinfo {pages} {3022} (\bibinfo {year}
  {2015})}\BibitemShut {NoStop}%
\bibitem [{\citenamefont {Schmidt}\ and\ \citenamefont
  {Thygesen}(2018)}]{schmidt2018benchmark}%
  \BibitemOpen
  \bibfield  {author} {\bibinfo {author} {\bibfnamefont {P.~S.}\ \bibnamefont
  {Schmidt}}\ and\ \bibinfo {author} {\bibfnamefont {K.~S.}\ \bibnamefont
  {Thygesen}},\ }\href@noop {} {\bibfield  {journal} {\bibinfo  {journal} {The
  Journal of Physical Chemistry C}\ }\textbf {\bibinfo {volume} {122}},\
  \bibinfo {pages} {4381} (\bibinfo {year} {2018})}\BibitemShut {NoStop}%
\bibitem [{\citenamefont {Furche}(2001)}]{furche2001molecular}%
  \BibitemOpen
  \bibfield  {author} {\bibinfo {author} {\bibfnamefont {F.}~\bibnamefont
  {Furche}},\ }\href@noop {} {\bibfield  {journal} {\bibinfo  {journal}
  {Physical Review B}\ }\textbf {\bibinfo {volume} {64}},\ \bibinfo {pages}
  {195120} (\bibinfo {year} {2001})}\BibitemShut {NoStop}%
\bibitem [{\citenamefont {Fuchs}\ and\ \citenamefont
  {Gonze}(2002)}]{fuchs2002accurate}%
  \BibitemOpen
  \bibfield  {author} {\bibinfo {author} {\bibfnamefont {M.}~\bibnamefont
  {Fuchs}}\ and\ \bibinfo {author} {\bibfnamefont {X.}~\bibnamefont {Gonze}},\
  }\href@noop {} {\bibfield  {journal} {\bibinfo  {journal} {Physical Review
  B}\ }\textbf {\bibinfo {volume} {65}},\ \bibinfo {pages} {235109} (\bibinfo
  {year} {2002})}\BibitemShut {NoStop}%
\bibitem [{\citenamefont {Ren}\ \emph {et~al.}(2012)\citenamefont {Ren},
  \citenamefont {Rinke}, \citenamefont {Joas},\ and\ \citenamefont
  {Scheffler}}]{ren2012random}%
  \BibitemOpen
  \bibfield  {author} {\bibinfo {author} {\bibfnamefont {X.}~\bibnamefont
  {Ren}}, \bibinfo {author} {\bibfnamefont {P.}~\bibnamefont {Rinke}}, \bibinfo
  {author} {\bibfnamefont {C.}~\bibnamefont {Joas}}, \ and\ \bibinfo {author}
  {\bibfnamefont {M.}~\bibnamefont {Scheffler}},\ }\href@noop {} {\bibfield
  {journal} {\bibinfo  {journal} {Journal of Materials Science}\ }\textbf
  {\bibinfo {volume} {47}},\ \bibinfo {pages} {7447} (\bibinfo {year}
  {2012})}\BibitemShut {NoStop}%
\bibitem [{\citenamefont {Marini}\ \emph {et~al.}(2006)\citenamefont {Marini},
  \citenamefont {Garc{\'\i}a-Gonz{\'a}lez},\ and\ \citenamefont
  {Rubio}}]{marini2006first}%
  \BibitemOpen
  \bibfield  {author} {\bibinfo {author} {\bibfnamefont {A.}~\bibnamefont
  {Marini}}, \bibinfo {author} {\bibfnamefont {P.}~\bibnamefont
  {Garc{\'\i}a-Gonz{\'a}lez}}, \ and\ \bibinfo {author} {\bibfnamefont
  {A.}~\bibnamefont {Rubio}},\ }\href@noop {} {\bibfield  {journal} {\bibinfo
  {journal} {Physical review letters}\ }\textbf {\bibinfo {volume} {96}},\
  \bibinfo {pages} {136404} (\bibinfo {year} {2006})}\BibitemShut {NoStop}%
\bibitem [{\citenamefont {Ras}\ \emph {et~al.}(2013)\citenamefont {Ras},
  \citenamefont {Louwerse}, \citenamefont {Mittelmeijer-Hazeleger},\ and\
  \citenamefont {Rothenberg}}]{ras2013predicting}%
  \BibitemOpen
  \bibfield  {author} {\bibinfo {author} {\bibfnamefont {E.-J.}\ \bibnamefont
  {Ras}}, \bibinfo {author} {\bibfnamefont {M.~J.}\ \bibnamefont {Louwerse}},
  \bibinfo {author} {\bibfnamefont {M.~C.}\ \bibnamefont
  {Mittelmeijer-Hazeleger}}, \ and\ \bibinfo {author} {\bibfnamefont
  {G.}~\bibnamefont {Rothenberg}},\ }\href@noop {} {\bibfield  {journal}
  {\bibinfo  {journal} {Physical Chemistry Chemical Physics}\ }\textbf
  {\bibinfo {volume} {15}},\ \bibinfo {pages} {4436} (\bibinfo {year}
  {2013})}\BibitemShut {NoStop}%
\bibitem [{\citenamefont {Takigawa}\ \emph {et~al.}(2016)\citenamefont
  {Takigawa}, \citenamefont {Shimizu}, \citenamefont {Tsuda},\ and\
  \citenamefont {Takakusagi}}]{takigawa2016machine}%
  \BibitemOpen
  \bibfield  {author} {\bibinfo {author} {\bibfnamefont {I.}~\bibnamefont
  {Takigawa}}, \bibinfo {author} {\bibfnamefont {K.-i.}\ \bibnamefont
  {Shimizu}}, \bibinfo {author} {\bibfnamefont {K.}~\bibnamefont {Tsuda}}, \
  and\ \bibinfo {author} {\bibfnamefont {S.}~\bibnamefont {Takakusagi}},\
  }\href@noop {} {\bibfield  {journal} {\bibinfo  {journal} {RSC advances}\
  }\textbf {\bibinfo {volume} {6}},\ \bibinfo {pages} {52587} (\bibinfo {year}
  {2016})}\BibitemShut {NoStop}%
\bibitem [{\citenamefont {Chowdhury}\ \emph {et~al.}(2018)\citenamefont
  {Chowdhury}, \citenamefont {Yang}, \citenamefont {Walker}, \citenamefont
  {Mamun}, \citenamefont {Heyden},\ and\ \citenamefont
  {Terejanu}}]{chowdhury2018prediction}%
  \BibitemOpen
  \bibfield  {author} {\bibinfo {author} {\bibfnamefont {A.~J.}\ \bibnamefont
  {Chowdhury}}, \bibinfo {author} {\bibfnamefont {W.}~\bibnamefont {Yang}},
  \bibinfo {author} {\bibfnamefont {E.}~\bibnamefont {Walker}}, \bibinfo
  {author} {\bibfnamefont {O.}~\bibnamefont {Mamun}}, \bibinfo {author}
  {\bibfnamefont {A.}~\bibnamefont {Heyden}}, \ and\ \bibinfo {author}
  {\bibfnamefont {G.~A.}\ \bibnamefont {Terejanu}},\ }\href@noop {} {\bibfield
  {journal} {\bibinfo  {journal} {The Journal of Physical Chemistry C}\
  }\textbf {\bibinfo {volume} {122}},\ \bibinfo {pages} {28142} (\bibinfo
  {year} {2018})}\BibitemShut {NoStop}%
\bibitem [{\citenamefont {Li}\ \emph {et~al.}(2017)\citenamefont {Li},
  \citenamefont {Wang}, \citenamefont {Chin}, \citenamefont {Achenie},\ and\
  \citenamefont {Xin}}]{li2017high}%
  \BibitemOpen
  \bibfield  {author} {\bibinfo {author} {\bibfnamefont {Z.}~\bibnamefont
  {Li}}, \bibinfo {author} {\bibfnamefont {S.}~\bibnamefont {Wang}}, \bibinfo
  {author} {\bibfnamefont {W.~S.}\ \bibnamefont {Chin}}, \bibinfo {author}
  {\bibfnamefont {L.~E.}\ \bibnamefont {Achenie}}, \ and\ \bibinfo {author}
  {\bibfnamefont {H.}~\bibnamefont {Xin}},\ }\href@noop {} {\bibfield
  {journal} {\bibinfo  {journal} {Journal of Materials Chemistry A}\ }\textbf
  {\bibinfo {volume} {5}},\ \bibinfo {pages} {24131} (\bibinfo {year}
  {2017})}\BibitemShut {NoStop}%
\bibitem [{\citenamefont {J{\"a}ger}\ \emph {et~al.}(2018)\citenamefont
  {J{\"a}ger}, \citenamefont {Morooka}, \citenamefont {Canova}, \citenamefont
  {Himanen},\ and\ \citenamefont {Foster}}]{jager2018machine}%
  \BibitemOpen
  \bibfield  {author} {\bibinfo {author} {\bibfnamefont {M.~O.}\ \bibnamefont
  {J{\"a}ger}}, \bibinfo {author} {\bibfnamefont {E.~V.}\ \bibnamefont
  {Morooka}}, \bibinfo {author} {\bibfnamefont {F.~F.}\ \bibnamefont {Canova}},
  \bibinfo {author} {\bibfnamefont {L.}~\bibnamefont {Himanen}}, \ and\
  \bibinfo {author} {\bibfnamefont {A.~S.}\ \bibnamefont {Foster}},\
  }\href@noop {} {\bibfield  {journal} {\bibinfo  {journal} {npj Computational
  Materials}\ }\textbf {\bibinfo {volume} {4}},\ \bibinfo {pages} {37}
  (\bibinfo {year} {2018})}\BibitemShut {NoStop}%
\bibitem [{\citenamefont {Zhang}\ and\ \citenamefont
  {Musgrave}(2007)}]{zhang2007comparison}%
  \BibitemOpen
  \bibfield  {author} {\bibinfo {author} {\bibfnamefont {G.}~\bibnamefont
  {Zhang}}\ and\ \bibinfo {author} {\bibfnamefont {C.~B.}\ \bibnamefont
  {Musgrave}},\ }\href@noop {} {\bibfield  {journal} {\bibinfo  {journal} {The
  journal of physical chemistry A}\ }\textbf {\bibinfo {volume} {111}},\
  \bibinfo {pages} {1554} (\bibinfo {year} {2007})}\BibitemShut {NoStop}%
\bibitem [{\citenamefont {Skriver}\ and\ \citenamefont
  {Rosengaard}(1992)}]{skriver1992surface}%
  \BibitemOpen
  \bibfield  {author} {\bibinfo {author} {\bibfnamefont {H.~L.}\ \bibnamefont
  {Skriver}}\ and\ \bibinfo {author} {\bibfnamefont {N.}~\bibnamefont
  {Rosengaard}},\ }\href@noop {} {\bibfield  {journal} {\bibinfo  {journal}
  {Physical Review B}\ }\textbf {\bibinfo {volume} {46}},\ \bibinfo {pages}
  {7157} (\bibinfo {year} {1992})}\BibitemShut {NoStop}%
\bibitem [{\citenamefont {Deb}\ \emph {et~al.}(1992)\citenamefont {Deb},
  \citenamefont {Singh},\ and\ \citenamefont {Sukumar}}]{deb1992universal}%
  \BibitemOpen
  \bibfield  {author} {\bibinfo {author} {\bibfnamefont {B.}~\bibnamefont
  {Deb}}, \bibinfo {author} {\bibfnamefont {R.}~\bibnamefont {Singh}}, \ and\
  \bibinfo {author} {\bibfnamefont {N.}~\bibnamefont {Sukumar}},\ }\href@noop
  {} {\bibfield  {journal} {\bibinfo  {journal} {Journal of Molecular
  Structure: THEOCHEM}\ }\textbf {\bibinfo {volume} {259}},\ \bibinfo {pages}
  {121} (\bibinfo {year} {1992})}\BibitemShut {NoStop}%
\bibitem [{\citenamefont {Takigawa}\ \emph {et~al.}(2018)\citenamefont
  {Takigawa}, \citenamefont {Shimizu}, \citenamefont {Tsuda},\ and\
  \citenamefont {Takakusagi}}]{takigawa2018machine}%
  \BibitemOpen
  \bibfield  {author} {\bibinfo {author} {\bibfnamefont {I.}~\bibnamefont
  {Takigawa}}, \bibinfo {author} {\bibfnamefont {K.-i.}\ \bibnamefont
  {Shimizu}}, \bibinfo {author} {\bibfnamefont {K.}~\bibnamefont {Tsuda}}, \
  and\ \bibinfo {author} {\bibfnamefont {S.}~\bibnamefont {Takakusagi}},\ }in\
  \href@noop {} {\emph {\bibinfo {booktitle} {Nanoinformatics}}}\ (\bibinfo
  {publisher} {Springer, Singapore},\ \bibinfo {year} {2018})\ pp.\ \bibinfo
  {pages} {45--64}\BibitemShut {NoStop}%
\bibitem [{\citenamefont {Pedregosa}\ \emph {et~al.}(2011)\citenamefont
  {Pedregosa}, \citenamefont {Varoquaux}, \citenamefont {Gramfort},
  \citenamefont {Michel}, \citenamefont {Thirion}, \citenamefont {Grisel},
  \citenamefont {Blondel}, \citenamefont {Prettenhofer}, \citenamefont {Weiss},
  \citenamefont {Dubourg} \emph {et~al.}}]{pedregosa2011scikit}%
  \BibitemOpen
  \bibfield  {author} {\bibinfo {author} {\bibfnamefont {F.}~\bibnamefont
  {Pedregosa}}, \bibinfo {author} {\bibfnamefont {G.}~\bibnamefont
  {Varoquaux}}, \bibinfo {author} {\bibfnamefont {A.}~\bibnamefont {Gramfort}},
  \bibinfo {author} {\bibfnamefont {V.}~\bibnamefont {Michel}}, \bibinfo
  {author} {\bibfnamefont {B.}~\bibnamefont {Thirion}}, \bibinfo {author}
  {\bibfnamefont {O.}~\bibnamefont {Grisel}}, \bibinfo {author} {\bibfnamefont
  {M.}~\bibnamefont {Blondel}}, \bibinfo {author} {\bibfnamefont
  {P.}~\bibnamefont {Prettenhofer}}, \bibinfo {author} {\bibfnamefont
  {R.}~\bibnamefont {Weiss}}, \bibinfo {author} {\bibfnamefont
  {V.}~\bibnamefont {Dubourg}},  \emph {et~al.},\ }\href@noop {} {\bibfield
  {journal} {\bibinfo  {journal} {Journal of machine learning research}\
  }\textbf {\bibinfo {volume} {12}},\ \bibinfo {pages} {2825} (\bibinfo {year}
  {2011})}\BibitemShut {NoStop}%
\bibitem [{\citenamefont {Sun}(2018)}]{sun2018kernel}%
  \BibitemOpen
  \bibfield  {author} {\bibinfo {author} {\bibfnamefont {F.}~\bibnamefont
  {Sun}},\ }\href@noop {} {\enquote {\bibinfo {title} {{Kernel Coherence
  Encoders, \textit{Master Thesis,Worcester Polytechnic Institute}}},}\ }
  (\bibinfo {year} {2018})\BibitemShut {NoStop}%
\bibitem [{\citenamefont {Friedman}\ \emph {et~al.}(2001)\citenamefont
  {Friedman}, \citenamefont {Hastie},\ and\ \citenamefont
  {Tibshirani}}]{friedman2001elements}%
  \BibitemOpen
  \bibfield  {author} {\bibinfo {author} {\bibfnamefont {J.}~\bibnamefont
  {Friedman}}, \bibinfo {author} {\bibfnamefont {T.}~\bibnamefont {Hastie}}, \
  and\ \bibinfo {author} {\bibfnamefont {R.}~\bibnamefont {Tibshirani}},\
  }\href@noop {} {\emph {\bibinfo {title} {The elements of statistical
  learning}}},\ Vol.~\bibinfo {volume} {1}\ (\bibinfo  {publisher} {Springer
  series in statistics New York},\ \bibinfo {year} {2001})\BibitemShut
  {NoStop}%
\bibitem [{\citenamefont {Liaw}\ \emph {et~al.}(2002)\citenamefont {Liaw},
  \citenamefont {Wiener} \emph {et~al.}}]{liaw2002classification}%
  \BibitemOpen
  \bibfield  {author} {\bibinfo {author} {\bibfnamefont {A.}~\bibnamefont
  {Liaw}}, \bibinfo {author} {\bibfnamefont {M.}~\bibnamefont {Wiener}},  \emph
  {et~al.},\ }\href@noop {} {\bibfield  {journal} {\bibinfo  {journal} {R
  news}\ }\textbf {\bibinfo {volume} {2}},\ \bibinfo {pages} {18} (\bibinfo
  {year} {2002})}\BibitemShut {NoStop}%
\end{thebibliography}%

\end{document}